\documentclass[aps,prl,twocolumn,groupedaddress,amsmath,amssymb,nofootinbib]{revtex4-2}
\usepackage[utf8]{inputenc}
\usepackage{graphicx,color}
\usepackage{caption}
\usepackage{subcaption}
\usepackage{amsmath}
\usepackage{epstopdf}
\usepackage{soul}

\usepackage{lipsum}
\newcommand{\D}{\mathrm{d}}
\newcommand{\e}{\mathrm{e}}

\newcommand{\be}{\begin{equation}}
\newcommand{\ee}{\end{equation}}
\newcommand{\bea}{\begin{eqnarray}}
\newcommand{\eea}{\end{eqnarray}}
\newcommand{\ba} {\begin{align} }
\newcommand{\ea} {\end{align} }

\newcommand{\kbt}{k_{\mathrm{B}}T}

\newcommand {\Na} {\text{Na}}
\newcommand {\K} {\text{K}}
\newcommand {\Cl} {\text{Cl}}

\newcommand{\tcl}{\tau_{\rm Cl}}
\newcommand{\sts}{\text{ss}}
\newcommand{\intra}{\text{i}}
\newcommand{\gt}{g_\text{tot}}

\begin{document}


\title{Separation of ionic timescales explains dynamics of cellular volume regulation}

\author{Ram M. Adar$^{1}$}
\email{radar@technion.ac.il}
\affiliation{$^1$ Department of Physics,Technion – Israel Institute of Technology, Haifa 32000, Israel
}

\begin{abstract}
Living cells actively regulate their volume in response to changes in the extra-cellular environment, such as osmolarity and chemo-attractant concentration. While the basic physical mechanisms of volume regulation are understood from the classic "pump-leak" model, it does not provide an explicit expression for the volume during dynamic regulation and can benefit from further insight into the volume dynamics. Here, we propose a simple explanation of volume dynamics in terms of two phases: fast volume adjustment to membrane potential, largely determined by $\Cl^-$ leakage, and slow potential adaptation after shock, constrained by $\Na^+$ leakage. The volume change may predominantly occur in either of these two phases, as we demonstrate for the scenarios of regulatory volume decrease and increase.  Our theoretical predictions are validated by two recent independent shock experiments: osmotic shocks in HeLa cells and neutrophil activation upon sudden exposure to chemoattractants. Our theory aims to elucidate cellular volume dynamics on the scale of tens of minutes in various biological contexts.
\end{abstract}


\maketitle
\section*{Introduction}
Cellular volume regulation plays a crucial role in maintaining homeostasis \cite{Hoffmann2009,Day2014,Cadart2019}. In animal cells, over periods of up to tens of minutes, this regulation is governed by water flux through the cell membrane and aquaporin channels, along with ion transport, commonly referred to as the "pump-leak" mechanism \cite{Tosteson1960,Kay2017,Jiang2013}. Ions diffuse passively (``leak'') through membrane-bound ion channels and are transported actively through energy-consuming pumps. The volume dynamics are then determined from the water flux, driven by differences in osmolarity across the cell membrane.

Homeostasis is challenged by external perturbations of the extra-cellular fluid, with osmolarity being the most frequently studied. Cellular response to such changes is tested in-vitro via osmotic shocks~\cite{Boudreault2004,Groulx2006,Fernandez2013,Roffay2021,VenkovaThesis}, where cells are exposed to a hypertonic/hypotonic medium, and their volume is measured over time. The cellular response typically comprises two phases \cite{Hoffmann2009,Cadart2019}: an initial passive volume decrease/increase within seconds, and subsequent adaptation and partial recovery of initial volume, which takes time of the order of minutes - ten minutes. This regulatory process is known as regulatory volume increase/decrease (RVI/RVD)~\cite{Lang1998}. It involves the activation of ion channels and transporters to modulate ion transport and cellular ionic concentrations. 

Perturbation may also occur in specific components of the extra-cellular environment. This is particularly pertinent for immune cells like neutrophils, which require rapid activation and response to pathogens. It was recently shown~\cite{Tamas} that neutrophils swell over approximately ten minutes upon exposure to chemoattractants 
due to regulation of ion transport.

In addition, cells change their volume in response to deformations. Three recent experiments~\cite{Guo2017,Xie2018,Venkova2022} have shown that cells lose volume during spreading. Spreading occurs within minutes, followed by volume loss with no significant delay. This effect was attributed to mechano-sensitive ion channels that open upon spreading and result in ion efflux~\cite{Xie2018,AdarPNAS,Venkova2022,AdarBJ}. The volume then decreases to retain osmotic pressure balance with the extra-cellular medium. 

 Active volume regulation has mostly been studied in the physiological context~\cite{Hoffmann2009}, including the identification and characterization of different ion channels and transporters that take part in the process. However, basic physical questions regarding cell volume dynamics are still open, namely: what sets its timescale? Why do deformations result in seemingly instantaneous volume loss, while regulatory volume dynamics can take around ten minutes? These questions are challenging to answer within the common pump-leak framework that requires numerically solving coupled, nonlinear rate equations for all ionic concentrations, without a clear physical intuition regarding the volume dynamics.
 
 This work introduces a new approach to address these questions, which focuses on the adjustment of cellular volume to membrane potential and the adaptation of membrane potential as part of RVD/RVI. The relevant timescales are derived from ion leakage times that are determined by ionic conductance and cell size.  Our theory is validated against recent independent experiments on various cell types and suggests new experiments to enhance our understanding of cellular volume regulation.

\section*{Methods}

The theoretical methods of the pump-leak model and linear response were used in order to predict the cellular volume dynamics, as detailed below.

\subsection*{Pump-leak model for cellular volume}
\label{sec1}

\begin{figure}[ht]
\centering
\includegraphics[width=0.9\columnwidth]{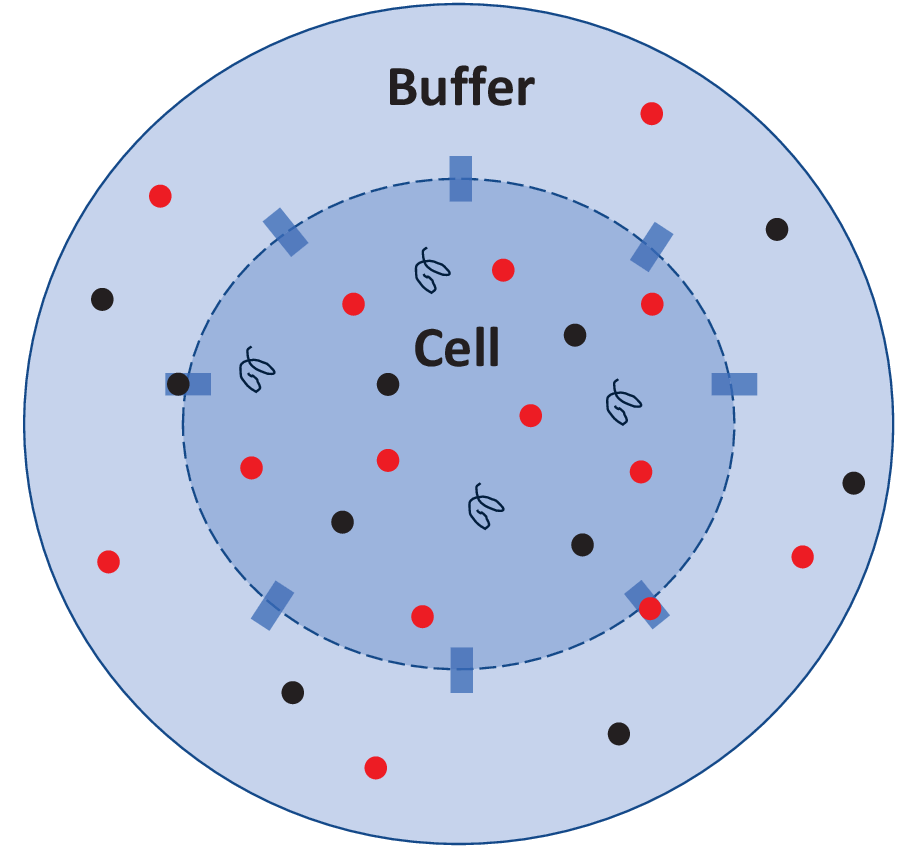}
\caption{(Color online) Sketch of a cell in suspension. The cell contains impermeant molecules with an average negative charge (black curly lines), anions (black particles) and cations (red particles). Ions exchange with the extracellular environment through ion channels and pumps embedded in the cell membrane, drawn as blue rectangles.}
\label{fig1}
\end{figure}

Our theory of cellular volume is based on the pump-leak model~\cite{Tosteson1960,Kay2017}. The cell has a volume $V$ and surface area $A$. The nucleus occupies a fixed volume fraction of the cell~\cite{Guo2017,Deviri2022,Romain2022} and can be coarse-grained~\cite{AdarPNAS}. The cellular content is modeled as an aqueous solution (see Fig.~\ref{fig1}). It consists mainly of water, ions that exchange with the (infinite) extracellular environment via ion channels and pumps, and impermeant mobile molecules, such as amino acids and proteins. The molar number of impermeant molecules $N$ evolves on long timescales~\cite{Cadart2019,Romain2022} that we are not concerned with here. We, therefore, consider $N$ to be fixed. 

The impermeant molecules have an average electric charge $-ze,$ where $-e$ is the electron charge and $z>0$ is the average valency in absolute value. This charge, together with active pumping, results in an asymmetry in ionic distributions and an electric potential difference across the cell membrane (membrane potential) $\psi<0$. The potential is considered to be homogeneous inside the cell~\cite{Donnan1924}, inducing no electric field.  

  The number of ions in the cell is determined by the passive ``leak'' down the electro-chemical potential gradient  and active pumping through the membrane, according to
\begin{align}
\label{eq1}
\frac{\D}{\D t}\left[n_\intra\left(V-V_d\right)\right]=\frac{A}{z_n F}\left[g_n\left(\frac{RT}{z_n F}\ln\frac{n_\e}{n_\intra}-\psi\right)+p_n\right].
\end{align}
Here, $n_\intra$ ($n_\e)$ is the intracellular (extracellular) molar concentration of ionic species $n$, $V-V_d$ is the effective cell volume, with $V_d$ the dry, non-aqueous volume due to macromolecules and intracellular membrane (obtained experimentally in the limit of infinite extracellular pressure). In addition, $R$ is the gas constant, $T$ is the absolute temperature, $F$ is Faraday's constant, and $z_n$ is the ionic valency. Note that $RT/F=\kbt/e.$ The passive transport rate is given by Ohm's law, in terms of the electric conductance $g_n$, where the term $RT\,\ln\left(n_\e/n_\intra\right)/ z_n F $ is sometimes referred to as the Nernst potential of ion $n$.  The pumping is given by the active current density $p_n$. Here we assume that the passive conductance and pump rates are fixed and not pH- or voltage-dependent. In particular, we limit our discussion to non-excitable cells, such as neutrophils and HeLa cells.

The classic pump-leak model~\cite{Keener,Kay2017} considers only the most abundant ions: K$^+$, Na$^+$, and Cl$^-$ and a single pump - the Na$^+$/K$^+$ exchanger. However, regulatory volume dynamics may also involve Cl$^-$ pumping~\cite{Delpire2018,Lang1998,Li2021,Boyd2022} as well as sodium-proton exchanger NHE~\cite{Li2021,Ni2024,Delpire2018,Lang1998}. This was specifically verified for neutrophil swelling upon exposure to chemo-attractants~\cite{Tamas}, which we discuss below. We thus allow a more general microscopic model, compared with the standard pump-leak one. However, as we focus on cellular volume and apply a coarse-grained approach, the details of this model will not be explicit in our theory, as evident below. 

Beyond ions, the cell also exchanges water with the extracellular environment, giving rise to volume dynamics, according to
\begin{align}
\label{eq2}
\frac{\D V}{\D t}=-L_w A\left(\Delta P-\Delta \Pi\right),
\end{align}
where $L_w$ is the membrane filtration coefficient (units of velocity per Pascal), while $\Delta P$ and $\Delta \Pi$ are the difference in hydrostatic and osmotic pressures, respectively, between the cell and the extracellular environment. The osmotic pressure is exerted by the ions and is approximated here by an ideal-gas pressure. Eq.~(\ref{eq2}) has a similar form to Eq.~(\ref{eq1}), where the volume is related to the number of water molecules and the difference between hydrostatic and osmotic pressures plays the role of the water chemical potential. Water transport, unlike ion transport, is only passive. 
\subsection*{Separation of time scales}
\label{sec2}
The pump-leak model is formulated in terms of the rate equations, Eqs. \eqref{eq1} and \eqref{eq2}. The common route is to solve them numerically [see, e.g., Refs. ~\cite{Kay2017,Li2021,Aminzare2024}], while possibly accounting for dynamics of specific transporters~\cite{Li2021,Ni2024}. We focus on volume dynamics, rather than ionic concentrations, and take a complementary route, based on separation of timescale. This approach provides both qualitative insight and quantitative predictions, as shown below. 

\begin{figure*}[ht]
\centering
\includegraphics[width=0.95\textwidth]{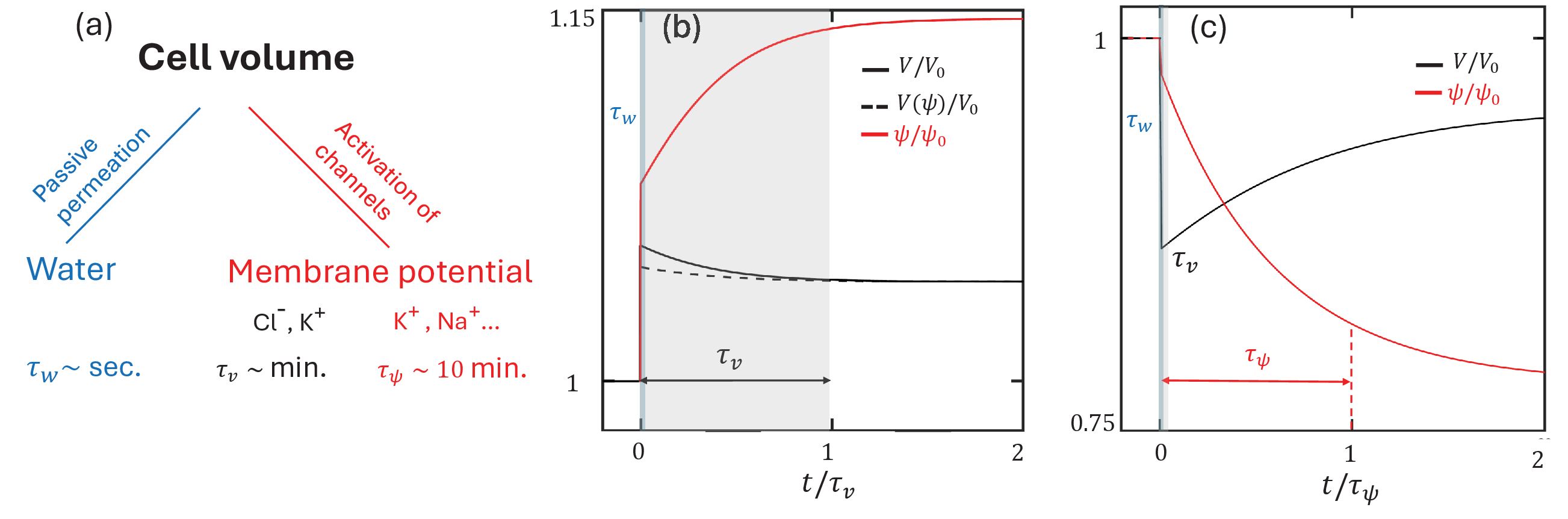}
\caption{(Color online) (a) Summary of the important physical mechanisms and timescales during cellular volume regulation up to tens of minutes. (b+c) Simple picture of cellular volume and membrane potential dynamics in response to hypotonic and hypertonic shocks (RVD and RVI, respectively),  where the subscript $0$ denotes initial value. The passive response occurs via water permeation during a typical time of $\tau_w\sim1$~sec, followed by volume adjustment to membrane potential over a time $\tau_v$, after which $V=V(\psi)$, as marked by the dashed black line. Changes in the membrane potential occur over a typical time $\tau_\psi$, limited by the slow leakage of Na$^+$. (b) RVD - most of the volume decreases during $\tau_v,$ while $\tau_\psi$ is much larger and not shown. (c) RVI - the volume increases during $\tau_\psi$ according to $V=V(\psi)$. The dashed black line overlaps with the solid one.}
\label{fig2}
\end{figure*}

Experimental evidence suggests at least three timescales involved in volume dynamics: (1) a time scale of the order of a second, characteristic of passive volume change in response to shocks, which is related to water flow. (2) A time scale of the order of seconds-minutes that allows for volume loss during spreading without evident delay. (3) A time scale of the order of minutes-ten minutes, characteristic of active regulatory volume increase/decrease (RVI/RVD). We relate these findings to three timescales of the pump-leak model  (see Fig.~\ref{fig2}): water permeation ($\tau_w$), volume adjustment to membrane potential ($\tau_v$), and membrane potential adaptation ($\tau_\psi$), which are well separated $\tau_w\ll\tau_v\ll\tau_\psi.$ This separation of timescales provides a simple framework for volume dynamics on the scale of tens of minutes: cellular volume adjusts quickly to the slowly adapting membrane potential. Next, we clarify how these timescales arise from the pump-leak model and focus on two limiting cases, where most of the volume change occurs during either $\tau_v$ or $\tau_\psi$. 
\\ \\
{\bf Water permeates quickly, making volume a fast variable.} The water timescale is found from Eq.~(\ref{eq2}) by considering the change in osmotic pressure due to water transport, while keeping the number of osmolytes in the cell fixed. This is given by $\tau_w=\left(V-V_d\right)/\left(L_w A \Pi_\intra\right)$, where $\Pi_\intra$ is the osmotic pressure in the cell. Inserting typical values of $\left(V-V_d\right)/A$ of order of $\mu$m and a bare membrane permeability $L_w\Pi_\intra$ of order of $10^2~\mu$m/sec~\cite{Milo2015} we find that $\tau_w<1$~sec. The value of $\tau_w$ implies that water permeates sufficiently fast to enforce mechanical equilibrium between the cell and the extracellular environment within seconds. This condition is well approximated by osmotic pressure balance, because of the low Laplace pressure that originates in cellular surface tension~\cite{Guo2017}. 

The cellular osmotic pressure $\Pi_\intra$ can be estimated from the concentration of osmolytes as $\Pi_\intra/R T=\left(1+z\right)N/\left(V-V_{d}\right)+2\Cl_\intra$, where the first term accounts for the impermeant molecules and their counterions, while the second term accounts for extraneous salt - intracellular Cl$^-$ together with its neutralizing K$^+$ and Na$^+$. Cl$^-$ ions have a unique role in determining the cellular volume, because they are not strictly required to neutralize the negative impermeant molecules. 

 Equating the cellular osmotic pressure with the extracellular pressure, $\Pi_\e=2\Cl_\e RT$, we find that~\cite{AdarBJ}:
  \begin{align}
	\label{eq4}
	v=\frac{V-V_{d}}{V_{m}-V_{d}} &=\frac{1}{1-\Cl_\intra/\Cl_\e}.
\end{align}
Here $v$ is a dimensionless volume, written in terms of the dry volume $V_d$ and  $V_{m}=V_{d}+\left(1+z\right)N/2\Cl_{\e}.$ The latter is the minimal possible volume in isotonic conditions, obtained for vanishing Cl$^-$ concentrations in cells with extremely negative membrane potential ($e \psi/\kbt\ll-1$). The volume thus follows the Cl$^-$ concentration, whose dynamics depend on the ionic timescales.

At steady state, the intracellular $\Cl^-$ concentration is given by $\Cl_\sts=\Cl_\e\exp\left[e\left(\psi_\sts-p_\Cl/g_\Cl\right)/\kbt\right].$  Together with Eq.~\eqref{eq4}, this provides a relation between cellular volume and membrane potential at steady-state. For simplicity, we focus hereafter on $p_\Cl=0,$ as in the standard pump-leak model. When $\Cl_\intra$ relaxes sufficiently  fast, as discussed below, it dynamically follows this expression, replacing $\psi_\sts$ by $\psi(t)$. In this case, the volume adjusts to the potential, $V=V[(\psi(t)]$, according to $v=1/\left[1-\exp\left(e\psi/\kbt\right)\right]$.
\\ \\
{\bf Conductance values determine ionic timescales and inter-ionic couplings.} The dynamics of ionic concentrations are mainly regulated by the ionic conductance values. To demonstrate this fact, we rewrite the pump-leak equations making use of two identities. First, we treat volume as a fast variable and insert [Eq.~\eqref{eq4}] in Eq.~\eqref{eq1}. Second, we write the ionic transport with respect to the steady state, given by the intracellular concentrations $n_\intra=n_\sts$ and membrane potential $\psi=\psi_\sts$. As the net flux of ions vanishes at steady state, we may subtract it from the equations. This leads to the following expression:
\begin{align}
\label{eq3a}
\frac{\D}{\D t}\left(\frac{n_\intra}{n_\sts}\frac{1-\Cl_\sts/\Cl_\e}{1-\Cl_\intra/\Cl_\e}\right)=\frac{1}{\tau_n}\left[\ln \frac{n_\sts}{n_\intra}-\frac{z_n e}{\kbt}\left(\psi-\psi_\sts\right)\right],
\end{align}
where we have denoted the ionic timescales
\begin{align}
\label{eq3}
\tau_n=\frac{F^2}{RT}\frac{n_\sts}{g_n}\frac{V_\sts-V_d}{A}\equiv\frac{1}{\beta_n}\frac{V_\sts -V_d}{A}.
\end{align}
The membrane potential acts as a Lagrange multiplier to ensure electroneutrality and is given by
\begin{equation}
    \label{eq3b}
    \frac{e}{\kbt}\left(\psi-\psi_\sts\right)=\frac{g_\K}{\gt}\ln\frac{\K_\sts}{\K_\intra}+\frac{g_\Na}{\gt}\ln\frac{\Na_\sts}{\Na_\intra}-\frac{g_\Cl}{\gt}\ln\frac{\Cl_\sts}{\Cl_\intra},
\end{equation}
where, $\gt=g_\K+g_\Na+g_\Cl$ is the total ionic conductance.

Equations~\eqref{eq3a}~-~\eqref{eq3b} provide an alternative formulation of the pump-leak model as an autonomous set of differential equations in terms of $n_\intra/n_\sts$. This form clearly demonstrates the two physical mechanisms that couple between the dynamics of the different ions. First, they occupy the same volume, and are thus coupled to $\Cl_\intra$. Second, the total change in ionic charge is zero (electroneutrality), as is ensured by the membrane potential $\psi$. This formulation also shows the different roles of active pumping and passive leak. Pumping only determines the steady state, while the leak also tunes the dynamics around steady state. This is because the latter is inherently coupled with the ionic concentrations, while the pump rates are not.
 
 The role of passive ionic conductances in the dynamics is twofold: they set the ionic timescales [Eq.~\eqref{eq3}] and the magnitude of inter-ionic couplings through the membrane potential [Eq.~\eqref{eq3b}]. 
 The timescale $\tau_n$ characterizes how long it will take all the ions of species $n$ to completely leak out of the cell (this does not actually happen). The timescale does not only depend on the conductance, but also on the intracellular concentration. It is faster for more dilute ions with high conductance. In addition, the timescale is smaller for small cells with large surface areas. In Eq.~\eqref{eq3} we have defined $\beta_n=g_n R T/ n_\sts F^2$ with units of velocity (similar to permeability).

The conductance $g_n$ is challenging to estimate. It is determined by the ion-channel composition on cellular scale, and can generally depend on cellular surface tension and membrane potential through stretch-activated and voltage-gated channels, respectively~\cite{Milo2015,Keener}. While we focus on non-excitable cells, ion conductance  has been studied most extensively in excitable cells. For this reason, we first demonstrate the separation of timescales based on the conductance values of squid giant axon at rest membrane potential~\cite{Hodgkin1952,Keener}: $g_\Na=0.01$ mS/cm$^2$, $g_\K=0.36$ mS/cm$^2$, and $g_\Cl=0.3$ mS/cm$^2$, where we have attributed the anionic current to Cl$^-$.  The typical intracellular ionic concentrations for these cells are~\cite{Keener} $\Na=50$\,mM, $\K=400$\,mM, and $\Cl=40$\,mM, leading to $\beta_\Na\approx 0.05~\mu$m/min., $\beta_\K\approx 0.1~\mu$m/min, and  $\beta_\Cl\approx 1~\mu$m/min. 

We do not expect these values to be universal across different cell types, but rather highlight the separation of timescales: the $\Na^+$ conductance is much lower than the other two, making $\Na^+$ transport less important around the rest membrane potential. In addition, $\Cl^-$ leakage is an order of magnitude faster than that of the cations. Note that our theory is not limited only to K$^+$, Na$^+$, and Cl$^-$ and may generally consist of additional timescales.
\\ \\
{\bf Timescales of volume adjustment ($\tau_v$) and potential adaptation ($\tau_\psi$).} As the model involves three ionic species with a single constraint (electroneutrality), the dynamics are described by two independent variables with typical rates. We propose the volume and membrane potential as these two variables, and divide the dynamics to two phases: (a) volume adjustment to the membrane potential during a timescale $\tau_v$ and (b) membrane potential adaptation during a timescale $\tau_\psi$, where the volume follows the potential according to $v=1/\left[1-\exp\left(e\psi/\kbt\right)\right]$. We distinguish between two limiting cases, where most of the volume dynamics occur during one of these phases. We show that the distinction between these two cases is determined by $\Na^+$ transport (more generally - by the less permeable cations) and we relate them to regulatory volume decrease/ increase (RVD/RVI).

\section*{Results}
\subsection*{Volume dynamics in response to shocks}
\label{ssec3}
We consider two types of shocks: osmotic shocks and sudden exposure to chemoattractants. Upon osmotic shock, cells are suddenly exposed to a hypotonic/hypertonic medium (each ionic concentration in the extracellular environment is modified by the same prefactor). The immediate passive response follows Ponder's relation~\cite{Ponder1930}, $\Pi_0\left(V_0-V_d\right)=\Pi_1\left(V_1-V_d\right)$, where $0$ and $1$ subscripts indicate values before and after the shock, respectively. This equality assumes that the number of osmolytes in the cell is unchanged and only water permeates through the membrane. 
Upon exposure to chemoattractants, the change in extracellular environment pressure is negligible and the passive response is related mostly to deformation. Neutrophils were shown to spread~\cite{Tamas}, leading to a volume loss (see Fig.~\ref{fig3}b). This effect has been measured in three additional, independent experiments on different cell types~\cite{Guo2017,Xie2018,Venkova2022} and was explained by ion transport via stretch-activated channels~\cite{Xie2018,AdarPNAS,Venkova2022,AdarBJ}. 

The passive volume response to both types of shocks is followed by active adaptation of ion transport rates: regulatory volume decrease (RVD) following hypotonic shock and regulatory volume increase (RVI) following hypertonic shock. Rather than considering specific ion transporters and their effects on each ionic species [see, e.g., Refs.~\cite{Li2021,Ni2024}], we apply a complementary coarse-grained approach, by focusing on cellular volume and membrane potential. 
\\ \\
{\bf Case 1: Volume dynamics during $\tau_v$ and RVD.} When $\Na_\intra$ changes are negligible around steady state (relevant for  $g_\Na/\gt\ll1$), volume dynamics are related only to $\K_\intra$ and $\Cl_\intra$. For each $\K^+$ ion that is transported, a $\Cl^-$ must be transported with it to maintain electroneutrality. There is, therefore, only one relevant timescale, $\tau_v$. The additional timescale, $\tau_\psi$ is related to further changes in membrane potential due to transport of $\Na^+$ ions, but it has a negligible effect on volume (see Fig.~\ref{fig2}b). This limit is relevant for RVD that relies on $\K^+$ and $\Cl^-$ transport, mainly via an increase in $g_\K$ and $g_\Cl$~\cite{Hoffmann2009,Delpire2018}. Consequently, the membrane potential becomes more negative (hyperpolarized) and the volume decreases.

This limit can be described mathematically by neglecting $g_\Na/\gt$. The set of Eqs.~\eqref{eq3a}-\eqref{eq3b} reduces to two coupled equations for $\K_\intra$ and $\Cl_\intra,$ while the latter directly relates to the volume via Eq.~\eqref{eq4}. This allows for the derivation of a single  equation for cellular volume (see SM Section 1). For small volume changes, it is given by 
\begin{align}
\label{eq7}
    v&=v_\sts-\Delta v\, \e^{-t/\tau_v},\nonumber\\
    \tau_v&=\left(1+\frac{g_\Cl}{g_\K}\right)\tau_\Cl,
    \end{align}
where $v_\sts$ is the dimensionless volume at steady state and $\Delta v=v_\sts-v(t=0)$ is the total volume change. There is a simple interpretation of $\tau_v$ in this limit: this is the time it will take all the intracellular $\Cl^-$ ions to leak together with their neutralizing $\K^+$ ions. For larger volume changes, a nonlinear equation can be derived for the volume. Both the linear and nonlinear versions show good agreement with a numerical solution of the full pump-leak model [Eq.~\eqref{eq1}], as shown in SM Section 1.
\\ \\
{\bf Case 2: Volume dynamics during $\tau_\psi$ and RVI.} When $\Na_\intra$ changes are considerable around steady state and have a significant effect on membrane potential, a different approach is necessary. 
We make use of the separation of timescales $\tau_\Cl\ll\tau_\K,\tau_\Na$, and consider that the $\Cl_\intra$ (and hence, the volume) quickly adjusts to the instantaneous membrane potential within a short time $\tau_v$, while most of the volume change is due to changes in membrane potential during $\tau_\psi$. This limit is relevant for RVI that relies on $\Na^+$ transport . Consequently, the membrane potential becomes less negative (depolarized) and the volume increases.

To capture this limit mathematically, we treat $\Cl_\intra$ as a fast variable with $\Cl^i(t)=\Cl_\e\exp\left[e\psi(t)/\kbt\right]$. The set of Eqs.~\eqref{eq3a}-\eqref{eq3b} reduces to two coupled equations for $\K_\intra$ and $\Na_\intra,$ which directly relate to the potential via Eq.~\eqref{eq3b}. For small potential changes,  we treat the potential dynamics within linear response and find that (see SM Section 2)
\begin{align}
\label{eq8}
    v&=\frac{1}{1-\exp\left[\frac{e}{\kbt}\left(\psi_\sts-\Delta\psi\,\e^{-t/\tau_\psi}\right)\right]},\nonumber\\
    \tau_\psi&=\frac{g_\K+g_\Na}{g_\K\,\tau_\K+g_\Na\,\tau_\Na}\tau_\K\,\tau_\Na,
    \end{align}
where $\Delta\psi=\psi_\sts-\psi(t=0)$ is the total change in membrane potential. The timescale of potential adaptation $\tau_\psi$ in this limit is due to the joint transport of $\Na^+$ and $\K^+$ ions and is constrained by the slower ion. For more general models that include additional ionic species [as is required, for example, to account for NHE exchangers as part of RVI~\cite{Hoffmann2009,Delpire2018,Ni2024}], $\tau_\psi$ is the slowest timescale of the linearized version of Eq.~\eqref{eq3a}. For larger potential changes, a nonlinear version can be derived for the dynamics. This theory also shows very good agreement with a numerical solution of the full pump-leak model [Eq.~\eqref{eq1}], as shown in SM Section 2.
\\ \\
{\bf Summary of volume response to shocks.} Our framework provides a qualitative explanation of volume dynamics upon shock. For example, we consider the response to a step-like hypertonic shock at time $t=0$ (see Fig. 2c).  Within a time $\tau_w,$ water flows out of the cell and leads to a passive shrinkage without changes in ion transport. This means that the volume scale $V_m$ in Eq.~\eqref{eq4} decreases due to the increased concentration of Cl$^-$ in the extracellular environment. Within our modeling, the shock is followed by a step-like activation of ion channels, which sets off RVI and changes the membrane potential according to Eq.~\eqref{eq3b}. The volume quickly adjusts within a time $\tau_v$ and continues to follow the instantaneous value of $\psi$. Then, within a time $\tau_\psi$, the potential adapts to a larger value (depolarization), set by the activation process [described by $\Delta\psi$ in Eq.~\eqref{eq8}]. 

Similarly, we can predict the response to osmolarity changes over a finite time $0\le t\le t_1$, corresponding to two consequent shocks; one at $t=0$ where the extracellular pressure changes from $\Pi_0$ to $\Pi_1>\Pi_0$ and a second shock at $t=t_1$, where the pressure changes back to $\Pi_0$. In this case, the active adaptation will result in a potential change of $\Delta\psi\left(1-\e^{-t_1/\tau_\psi}\right)$, while the volume scale $V_m$ will go back to its original value. Therefore, we expect the volume to "overshoot" at $t=t_1$ to a larger volume than the original volume at $t=t_0$, as long as $t_1$ is not too small compared to $\tau_\psi$. The same arguments hold for the case of hypotonic shock and RVD, with $\tau_v$ being the relevant timescale. In both cases, subsequent volume regulation may follow.

Beyond this qualitative understanding of the volume dynamics, our theory provides quantitative predictions for different shock experiments, as demonstrated next. 

\subsection*{Comparison with shock experiments}
\label{sec3}
We compare our predictions with two recent, independent shock experiments with different kinds of shocks on two cell types. The volume measurement in both experiments was carried out using Fluorescence Exclusion Microscopy~\cite{Bottier2011,Cadart2017}.
\\

\begin{figure*}[ht]
\centering
\includegraphics[width=0.95\textwidth]{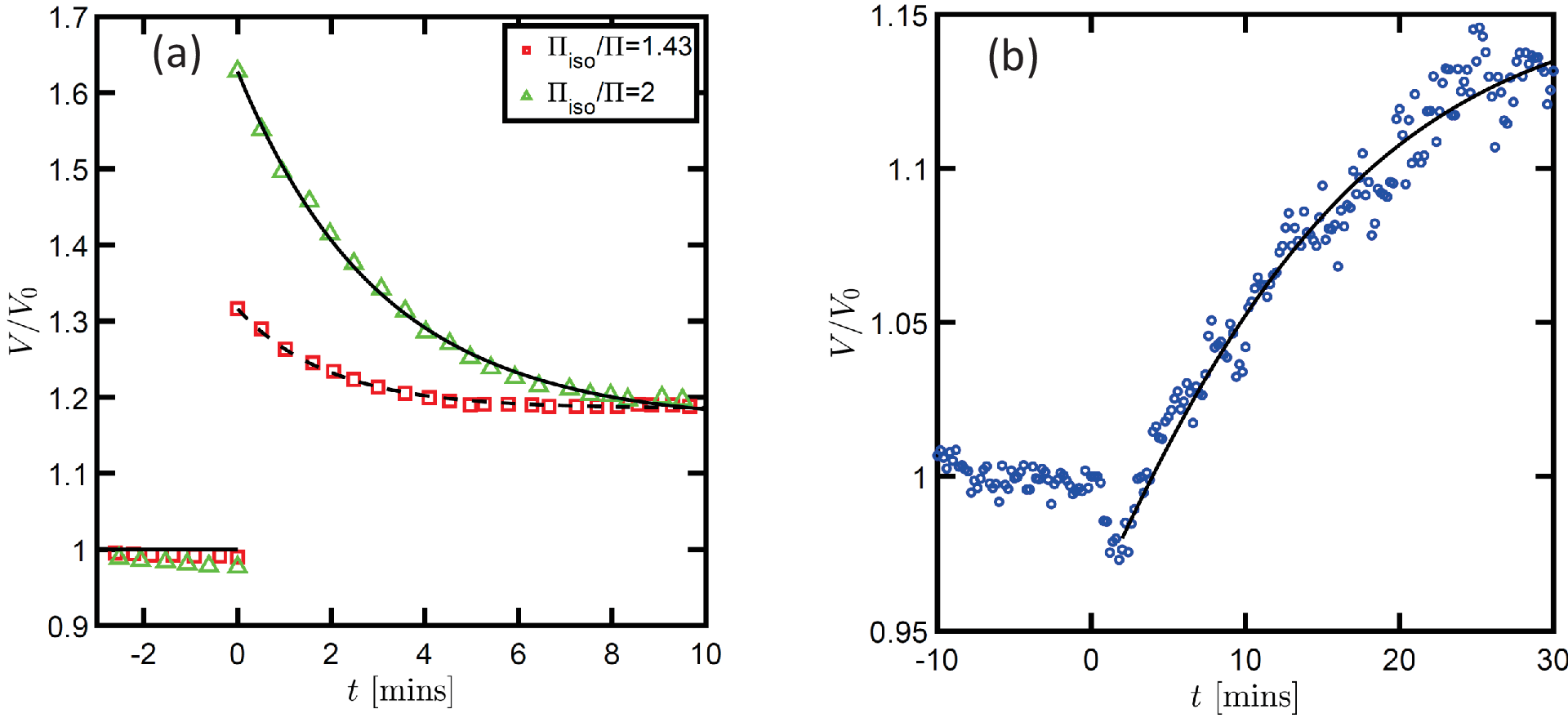}
\caption{(Color online) (a) Normalized volume of HeLa cells as a function of time for two hypotonic shock experiments~\cite{VenkovaThesis}, $\Pi/\Pi_{\rm iso}=1.43, 2$ (blue  red squares and green triangles, respectively). The fitted values are $\tau_v=1.9, 3.1$ minutes and  $\Delta v=-0.21, -0.53$  (dashed line and solid line, respectively). (b) Normalized volume of neutrophil cells as a function of time upon exposure to chemoattractnats. The experimental data~\cite{Tamas} is plotted in blue circles and the fit is plotted as a solid black line for $\tau_\psi=11.4$ minutes and $e\Delta \psi/\kbt=0.30$. The values $V_d/V_0=0.29$ and $V_m/V_d=2.5$ were used for all the data sets in (a) and (b). }
\label{fig3}
\end{figure*}

{\bf Osmotic shocks in HeLa cells (RVD).} 
First, we consider an osmotic shock experiment in HeLa cells~\cite{VenkovaThesis}. The cells were exposed to a hypotonic medium at time $t=0$, and their volume was measured over time (see Fig.~\ref{fig3}a). The initial passive volume response was considered as instantaneous and was predicted according to Ponder's relation with $V_d/V_0=0.29,$ as was extracted in Ref.~\cite{VenkovaThesis} by fitting the passive volume response for different extracellular environment pressures. The active volume response was predicted using Eq.~(\ref{eq7}) with $V_m/V_d=2.5$ (corresponding to an average volume of about $20$~nm$^3$ per impermeant osmolyte), while $\tau_v$ and $\Delta v$ serve as fit parameters.
Our predictions fit well with the experimental data, as evident from Fig.~\ref{fig3}a.  The shock $\Pi/\Pi_{\rm iso}=2$ (where $\Pi_{\rm iso}$ is the osmotic pressure in isotonic condition) was best fitted by $\tau_v=3.1$ minutes and $\Delta v=-0.53$ (solid line), whereas the shock $\Pi/\Pi_{\rm iso}=1.43$  was best fitted by $\tau_v=1.9$ minutes and $\Delta v=-0.21$ (dashed line). 
\\\\
{\bf Neutrophil exposure to chemo-attractants (RVI).} Second, we consider the experiments of Ref.~\cite{Tamas}, where  neutrophils where exposed to chemo-attractants and their volume was measured over time. Upon exposure, neuotrophils become activated. They spread on their underlying surface on the scale of a minute and lose volume, similarly to the effect measured in Refs.~\cite{Guo2017,Xie2018,Venkova2022}. This behavior can be explained by stretch-activated channels~\cite{Xie2018,AdarPNAS,Venkova2022,AdarBJ}. The fast volume loss is justified in this case by the short $\tau_v$ timescale underlying our theory of RVI.

After spreading, neutrophils swell within about ten minutes, due to the activation of ion transporters [the effect vanishes upon knockdown of sodium-proton exchanger NHE1~\cite{Tamas}].  We focus on the volume dynamics during activation and compare the experimental data with our theoretical prediction for RVI dynamics, based on Eq.~\eqref{eq8}, using $\Delta \psi$ and $\tau_\psi$ as fit parameters.
Our predictions fit very well with the experimental data, as is shown in Fig.~\ref{fig3}b. For simplicity, we used the same values of $V_d/V_0=0.29$ and $V_m/V_d=2.5$. The best-fitted parameters were $e\Delta \psi/\kbt=0.30$ and $\tau_\psi=11.4$ minutes.  Note that the timescale here is larger than that fitted for the hypotonic shocks. This is consistent with the underlying dynamics of RVI that involve the slower $\Na^+$, rather than just $\K^+$, as in RVD. In addition, the experiments involve different cell types (HeLa and neutrophils), which may have different ionic timescales.

\section*{Discussion}
\label{sec4}
Our work presents a straightforward explanation of regulatory volume dynamics in response to shocks: cellular volume  adjusts to the membrane potential during a typical volume timescale $\tau_v$,  while membrane potential adapts more slowly during the potential timescale $\tau_\psi$, constrained by slow cation leakage. We highlight two limiting cases, where most of the volume dynamics are during either $\tau_v$ or $\tau_\psi$ and relate them to RVD and RVI, respectively. Furthermore, the possible short $\tau_v$ timescale thanks to fast $\Cl^-$ leakage explains how volume loss may occur without evident delay during spreading. 

We provide several predictions that distinguish between RVD and RVI. For example, we estimate different RVD and RVI timescales, related to the difference between $\Cl^-$ and $\Na^+$ conductance. Furthermore, we expect cell volume to follow the potential without evident delay only during RVI (more generally, the volume should follow $\psi-p_\Cl/g_\Cl$).  To experimentally validate our interpretation and predictions, simultaneous measurement of the membrane potential and cellular volume using fluorescence lifetime imaging~\cite{Lazzari2019} can be employed. 

The adjustment of cellular volume to membrane potential is largely facilitated by the short Cl$^-$ leakage time, $\tcl$. We estimate this to be an order of magnitude shorter than the adaptation time of membrane potential $\tau_\psi$, which is related to K$^+$ and Na$^+$ leakage times. This critical role of Cl$^-$ dynamics in volume regulation is often overlooked in the pump-leak model, which focuses primarily on Na$^+$ and K$^+$ due to the significant role of the Na$^+$/K$^+$ exchanger. Our analysis demonstrates that active pumping determines the steady state, while the leakage - and especially that of $\Cl^-$ - sets the dynamics.

We note that the ionic timescales are defined by the new steady state, after the activation of channels, rather than the reference state before shock. This factor may be important. For example, Ehrlich ascites tumor cells~\cite{Lambert1989} exhibit low Cl$^-$ conductance in their reference state, $g_\Cl\approx0.06 g_\K$. However, upon swelling, the conductance values change drastically such that $g_\Cl\approx2g_\K$. Furthermore, the separation of timescales that we suggest is not necessarily universal. To gain a better understanding of volume regulation dynamics, it is essential to measure and characterize the conductance values of the various ions across a wider range of cell types, as well as their regulatory volume dynamics. 

The dynamics of volume regulation are complex due to the numerous cellular degrees of freedom and nonlinear couplings between them, as demonstrated by Eq.~\eqref{eq1}. The aim of this work is to simplify this dynamical system as much as possible and provide physical intuition. To this end, we have focused on two degrees of freedom: membrane potential and cell volume, and two corresponding timescales:  $\tau_v$ and $\tau_\psi$. One reason for the success of our approach is the absence of additional timescales in the external perturbation. We limited this work to cellular responses to shocks, as opposed to perturbations with intrinsic timescales, such as oscillatory osmolarity changes. Perturbations with a period that is much smaller than any of $\tau_v,\,\tau_\psi$ will require the consideration of shorter timescales. 

Additional timescales may also arise from the cellular response, including those related to Ca$^{2+}$ transport, and possible regulation of active pumping rates.  Moreover, the regulatory modification of ion transport rates, which was modeled here as a simple step function,  may have intrinsic timescales. For example, it was recently shown that regulatory volume increase can be triggered by phase separation of WNK kinases~\cite{Boyd2022} upon cell shrinkage. While this process usually occurs within seconds, modifications of the disordered domain of the kinase affected the phase-separation dynamics and introduced an additional timescale of about five minutes to the volume response. Another example is the K-Cl co-transporter, sometimes involved in RVD, which is known to exhibit a lag phase in its activation dynamics~\cite{Delpire2018,Jennings1990}. Overall, our framework is useful as long as these additional timescales can be neglected. 

Finally, we emphasize that our theory is formulated within a mean-field framework and describes only average dynamics, while cells exhibit variability in volume regulation. This variability was recently demonstrated in HeLa cells~\cite{VenkovaThesis}, where some cells displayed no RVD in response to hypotonic shock. In our theory, the magnitude of the regulatory response is modeled phenomenologically by the parameters $\Delta v$ and $\Delta\psi$. In the future, we propose to experimentally investigate the distribution of cellular volume dynamics and, based on the findings, develop a statistical-mechanical model of ion transporter activation and consequent volume dynamics. 
\\
{\it Acknowledgments.} We thank L. Venkova, A. R. Kay, and M. Piel for useful discussions and K. Keren, E. Braun, and S. A. Safran for a critical reading of the manuscript.

\bibliography{Refs}


\begin{widetext}
\section*{Supporting Material}

\subsection{1. Derivation of the volume dynamics during $\tau_v$ (RVD) }
Here we derive Eq.~(7) of the main text as well as its nonlinear generalization. Assuming that the relative $\Na^+$ conductance is negligible ($g_\Na/g_\text{tot}\ll1$), the difference in ionic content with respect to steady-state can be written only in terms of $\K^+$ and $\Cl^-$. In this limit, after substituting the expression for the membrane potential [Eq. (6) of the main text], Eq.(4) of the main text reads
\begin{align}
\label{eqsa1}
\frac{\D}{\D t}\left(\frac{\K_\intra}{1-\Cl_\intra/\Cl_\e}\right)&=\frac{RT}{F^2}\frac{A}{V_m-V_d}\frac{g_\K\,g_\Cl}{g_\K+g_\Cl}\ln\frac{\K_\sts \Cl_\sts}{K_\intra\Cl_\intra},\nonumber\\
\frac{\D}{\D t}\left(\frac{\Cl_\intra}{1-\Cl_\intra/\Cl_\e}\right)&=\frac{RT}{F^2}\frac{A}{V_m-V_d}\frac{g_\K\,g_\Cl}{g_\K+g_\Cl}\ln\frac{\K_\sts \Cl_\sts}{K_\intra\Cl_\intra}.
\end{align}

Eq.~\eqref{eqsa1} demonstrates that the rate of change in the total number of $\K^+$ and $\Cl^-$ ions is exactly the same to maintain electroneutrality (in the limit of negligible change in $\Na^+$ ions). Integrating over time, we find that the change in the number of ions is the same, allowing us to write the intracellular $\K^+$ concentration in terms of the $\Cl^-$ concentration:
\begin{equation}
\label{eqsa2}
\frac{\K_\intra}{1-\Cl_\intra/\Cl_\e}-\frac{\K_\sts}{1-\Cl_\sts/\Cl_\e}=\frac{\Cl_\intra}{1-\Cl_\intra/\Cl_\e}-\frac{\Cl_\sts}{1-\Cl_\sts/\Cl_\e}.
\end{equation}
Inserting this identity in the second row of Eq.~\eqref{eqsa1}, yields an autonomous equation for $\Cl_\intra$.

We make use of the relation between the $\Cl^-$ concentration and dimensionless cell volume, $v=1/\left(1-\Cl_\intra/\Cl_\e\right)$, to write an equation for the dimensionless volume
\begin{equation}
\label{eqsa3}
\frac{\D v}{\D t}=-\frac{RT}{F^2}\frac{A}{V_m-V_d}\frac{1}{\Cl_\e}\frac{g_\K\,g_\Cl}{g_\K+g_\Cl}\ln\left(\frac{v-1}{v}\frac{v_\sts}{v_\sts-1}\frac{\Cl_\e}{\K_\sts}\left[1-\frac{v_\sts}{v}\left(1-\frac{\K_\sts}{\Cl_\e}\right)\right]\,\right).
\end{equation}

The timescale $\tau_v$ is defined by the linearized equation, $\D v/\D t=-\left(v-v_\sts\right)/\tau_v$. We linearize the equation and find that 
\begin{equation}
\label{eqsa4}
\frac{\D v}{\D t}\approx-\frac{RT}{F^2}\frac{A}{V_m-V_d}\frac{1}{\Cl_\e}\frac{g_\K\,g_\Cl}{g_\K+g_\Cl}\times\left[\Cl_\e\left(\frac{1}{\K_\sts}+\frac{1}{\Cl_\sts}\right)-2\right]\frac{v-v_\sts}{v_\sts}.
\end{equation}
We further approximate the term in the brackets 
\begin{equation}
\label{eqsa4}
\Cl_\e\left(\frac{1}{\K_\sts}+\frac{1}{\Cl_\sts}\right)-2\approx\frac{\Cl_\e}{\Cl_\sts}=\frac{v_\sts}{v_\sts-1}.
\end{equation}
Overall, this yields:
\begin{align}
\label{eqsa5}
    \frac{\D v}{\D t}&=-\frac{v_\sts-1}{\tau_v}\ln\left(\frac{v-1}{v}\frac{v_\sts}{v_\sts-1}\frac{\Cl_\e}{\K_\sts}\left[1-\frac{v_\sts}{v}\left(1-\frac{\K_\sts}{\Cl_\e}\right)\right]\,\right),\nonumber\\
    \tau_v&=\frac{F^2}{RT}\,\frac{V_\sts-V_d}{A}\,\Cl_\sts\,\frac{g_\K+g_\Cl}{g_\K\, g_\Cl}=\left(1+\frac{g_\Cl}{g_\K}\right)\tau_\Cl.
    \end{align}
Linearization of $v$ around $v_\sts$ results in $\D v/\D t=-\left(v-v_\sts\right)/\tau_v$, whose solution is Eq. (7) of the main text,
\begin{equation}
\label{eqsa6}
v=v_\sts-\Delta v\, \e^{-t/\tau_v}.
\end{equation}

This simplified theory of volume dynamics [Eqs.~\eqref{eqsa5}-\eqref{eqsa6}] shows very good agreement with a numerical solution of the full pump-leak model [Eqs.~(1)-(3) of the main text], as shown in Fig.~\ref{figSM1}a. The linear Eq.~\eqref{eqsa6} works well especially close to steady state, as is expected, while the nonlinear Eq.~\eqref{eqsa5} works very well throughout the entire dynamics. Furthermore, comparing the volume with $v(\psi)=1/\left[1-\exp\left(e\psi/\kbt\right)\right]$ (dahsed black line) demonstrates that the volume change indeed occurs during the adjustment of the volume to the potential with a typical timescale $\tau_v$.

In order to solve the pump-leak model numerically, we have inserted $v=1/\left(1-\Cl_\intra/\Cl_\e\right)$ in Eq. (1) of the main text and integrated the equations numerically. We consider the following extracellular concentrations (in mM): $\K_\e=145$, $\Na_\e=5,$ and $\Cl_\e=150.$  As an initial condition, we have considered the steady state corresponding to the conductance values $g_\K=0.36g$, $g_\Na=0.01g$ and $g_\Cl=0.3g,$ where $g$ is an intrinsic conductance scale that enters $\tau_v$, with the pumping rates $p_\K=2p,$ $p_\Na=-3p$ and $p=0.08g$. The regulatory dynamics are initiated at $t=0$ by assuming a 40-fold increase in $g_\K$ and $g_\Cl$, while keeping $g_\Na$ fixed. The membrane potential is determined by Eq. (6) of the main text.

\begin{figure}[ht]
\centering
\includegraphics[width=0.9\columnwidth]{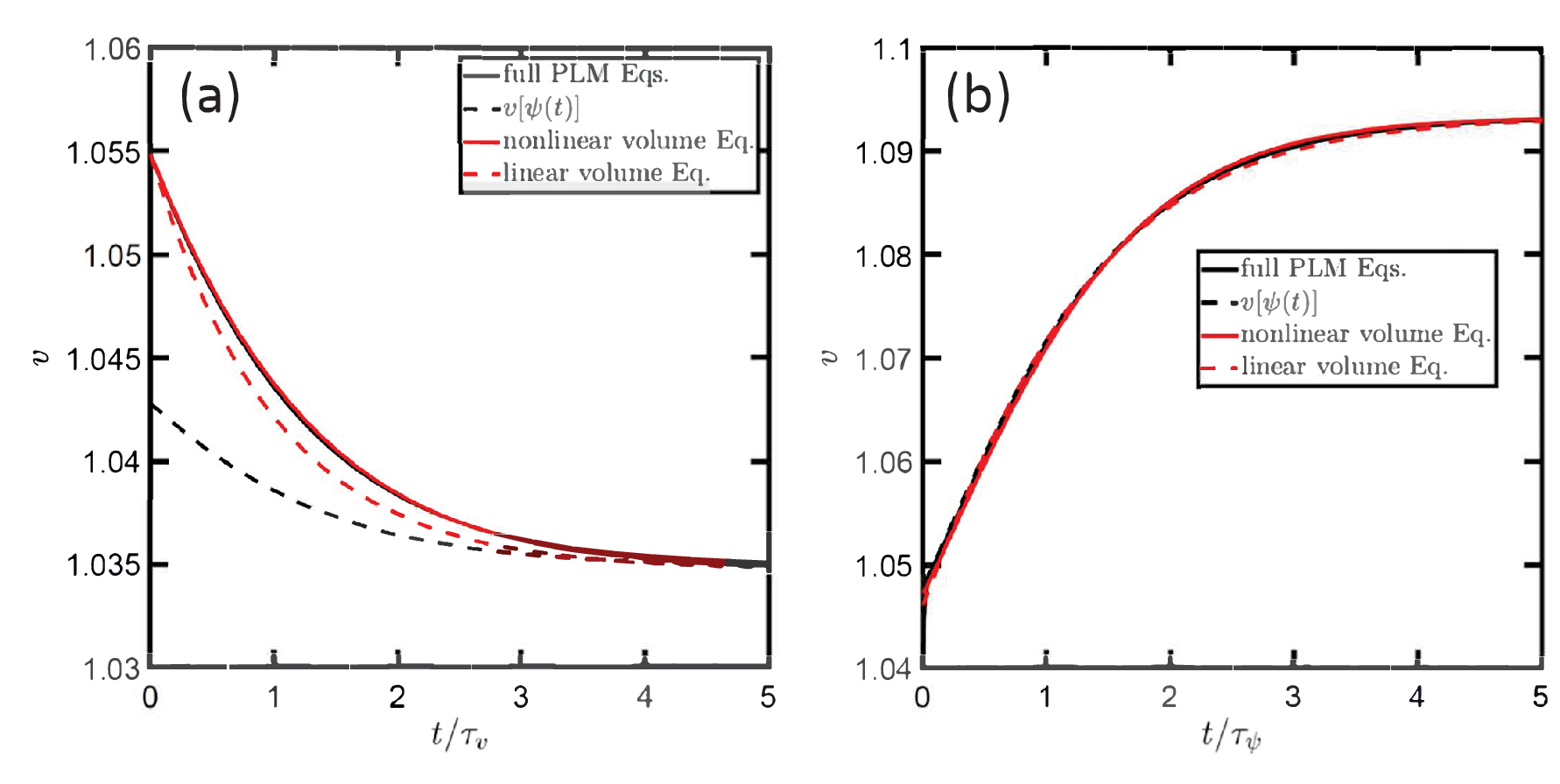}
\caption{(Color online) Predictions regarding regulatory volume dynamics and comparison with numerical solutions of the pump-leak model. (a) Predictions for RVD, based on the nonlinear theory [Eq.~\eqref{eqsa5}, solid red] and linear theory [Eq.~\eqref{eqsa6}, dashed red], compared with numerical solution of the pump-leak model [Eqs. (1)-(3) of the main text, solid black] and the potential-adjusted volume $v(\psi)=1/\left[1-\exp\left(e\psi/\kbt\right)\right]$ (dashed black). (b) Predictions for RVI, based on the nonlinear theory [Eqs.~\eqref{eqsb8} and \eqref{eqsb10}, solid red] and linear theory [Eq.~\eqref{eqsb6a}, dashed red], compared with numerical solution of the pump-leak model (solid black) and the potential-adjusted volume (dashed black). }
\label{figSM1}
\end{figure}
\newpage
\subsection{2. Derivation of the volume dynamics during $\tau_\psi$ (RVI)}
Here we derive Eq.~(8) of the main text as well as its nonlinear generalization. Treating $\Cl_\intra$ as a fast variable, we find that $\Cl_\intra(t)=\Cl_\e \exp\left[e\psi(t)/\kbt\right]$ and, accordingly, 
\begin{align}
\label{eqsb1}
v(t)&=\frac{1}{1-\exp\left[e\psi(t)/\kbt\right]}.
\end{align}
The dynamic equations can then be written in terms of $\Na_\intra$ and $\K_\intra,$ which are coupled by the membrane potential
\begin{equation}
\label{eqsb2}
\frac{e}{\kbt}\left(\psi-\psi_\sts\right)=\frac{g_\K}{g_\K+g_\Na}\ln\frac{\K_\sts}{\K_\intra}+\frac{g_\Na}{g_\K+g_\Na}\ln\frac{\Na_\sts}{\Na_\intra}.
\end{equation}

Substituting this expression in Eq. (4) of the main text yields coupled equations for the cations
\begin{align}
\label{eqsb3}
\frac{\D}{\D t}\left(\frac{\K_\intra}{\K_\sts} \frac{v}{v_\sts}\right)&=\,\,\frac{1}{\tau_\K}\frac{g_\Na}{g_\K+g_\Na}\left(\ln\frac{\K_\sts}{\K_\intra}-\ln\frac{\Na_\sts}{\Na_\intra }\right),\nonumber\\
\frac{\D}{\D t}\left(\frac{\Na_\intra}{\Na_\sts} \frac{v}{v_\sts}\right)&=\frac{1}{\tau_\Na}\frac{g_\K}{g_\K+g_\Na}\left(\ln\frac{\Na_\sts}{\Na_\intra }-\ln\frac{\K_\sts}{\K_\intra}\right).
\end{align}

We proceed in two possible ways. First, for small potential changes, we treat $\psi(t)$ within linear response. Second, we suggest a full, nonlinear treatment of the dynamics.

{\bf Linear response.} We consider small changes in the ionic concentrations, $\K_\intra\approx\K_\sts\left(1+\delta\K\right)$, $\Na_\intra\approx\Na_\sts\left(1+\delta\Na\right)$ for $\delta\K,\delta\Na\ll 1$. The  volume is modified accordingly, $v\approx v_\sts\left(1+\delta v\right).$ We expand Eq.~\eqref{eqsb3}  to linear order and find that
\begin{align}
    \label{eqsb5}
\frac{\D}{\D t}\delta\K+\frac{\D}{\D t}\delta v&=-\frac{1}{\tau_\K}\frac{g_\Na}{g_\K+g_\Na}\left(\delta\K-\delta\Na\right),\nonumber\\
\frac{\D}{\D t}\delta\Na+\frac{\D}{\D t}\delta v&=-\frac{1}{\tau_\Na}\frac{g_\K}{g_\K+g_\Na}\left(\delta\Na-\delta\K\right).
\end{align}
Here $\delta v$ depends on $\delta\K$ and $\delta\Na$, but this dependence is not important for the analysis, as shown next. 

We insert $\delta\Na, \delta\K\sim\exp\left(st\right)$,  where $s\le0$ is the typical relaxation rate, and find the values of $s$ that solve the linear system. As the right-hand-side of Eq.~(\ref{eqsb5}) depends only on $\delta\K-\delta\Na$, it is evident that the determinant of linear system vanishes for $s=0$, making it an eigenvalue. This mode is enforced by the membrane potential (electroneutrality). The second mode is found by subtracting the two equations. The linear combination $\delta\K-\delta\Na$ decays with a rate $s=1/\tau_\psi$ with
\begin{align}
    \label{eqsb6}
\tau_\psi&=\frac{g_\K+g_\Na}{g_\K\,\tau_\K+g_\Na\,\tau_\Na}\tau_\K\,\tau_\Na.
\end{align}
In a more general model $\tau_\psi$ is the longest timescale of the ion dynamics and it determines the duration of membrane potential adaptation. The volume immediately adjusts to the potential, because of fast Cl$^-$ transport. As $\psi$ changes due to a combination of $\delta \K$ and $\delta\Na$, it should evolve with the same time scale, according to
\begin{align}
    \label{eqsb6a}
\psi&=\psi_\sts-\Delta\psi\exp\left(-t/\tau_\psi\right).
\end{align}
\\
{\bf Nonlinear theory.} As an alternative approach, we rewrite Eq.~\eqref{eqsb3} in terms of the number of intracellular cations $\sim\K_\intra v,\Na_\intra v$. Rearranging the terms and inserting the expressions for $\tau_\K$ and $\tau_\Na$, we find that
\begin{align}
\label{eqsb7}
\frac{\D}{\D t}\left(\K_\intra v\right)&=\frac{RT}{F^2}\frac{A}{V_m-V_d}\frac{g_\K \,g_\Na}{g_\K +g_\Na}\left(\ln\frac{\K_\sts v_\sts}{\K_\intra v}-\ln\frac{\Na_\sts v_\sts}{\Na_\intra v}\right),\nonumber\\
\frac{\D}{\D t}\left(\Na_\intra v\right)&=\frac{RT}{F^2}\frac{A}{V_m-V_d}\frac{g_\K \,g_\Na}{g_\K +g_\Na}\left(\ln\frac{\Na_\sts v_\sts}{\Na_\intra v}-\ln\frac{\K_\sts v_\sts}{\K_\intra v}\right).
\end{align}
Note that we have added and subtracted $\ln\left(v_\sts/v\right)$ in the parenthesis on the right-hand-side. 

We find that the number of cations $\sim v\left(\K_\intra+\Na_\intra\right)$ is conserved in this limit. Therefore, we can use the identity $\Na_\intra v-\Na_\sts v_\sts=\K_\sts v_\sts-\K_\intra v$. This leads to the following equation for $x=\K_\intra v$:
\begin{align}
\label{eqsb8}
   \frac{\D x }{\D t}&=-\frac{v_\sts}{\tau_\psi}\frac{\K_\sts \,\Na_\sts}{\K_\sts +\Na_\sts}\ln\left[\frac{\left(\Na_\sts /\K_\sts\right)\,x}{\left(\Na_\sts+\K_\sts\right)v_\sts-x}\right].
    \end{align}

Finally, we relate $\K_\intra v$ to the membrane potential and $v$, according to 
\begin{align}
\label{eqsb9}
\frac{e}{\kbt}\left(\psi-\psi_\sts\right)&=\frac{g_\K}{g_\K+g_\Na}\ln\frac{\K_\sts v_\sts}{\K_\intra v}+\frac{g_\Na}{g_\K+g_\Na}\ln\frac{\Na_\sts v_\sts}{\Na_\intra v}+\ln\frac{v}{v_\sts}.
\end{align}
Inserting this in the expression for $v$ [Eq.~\eqref{eqsb1}] leads to 
\begin{equation}
\label{eqsb10}
v=\frac{1}{1-\frac{v}{v_\sts}\frac{v_\sts -1}{v_\sts}\left(\frac{\K_\sts\,v_\sts}{x}\right)^\frac{g_\K}{g_\K+g_\Na}\left[\frac{\Na_\sts\,v_\sts}{\left(\Na_\sts+\K_\sts\right)v_\sts-x}\right]^\frac{g_\Na}{g_\K+g_\Na}}\equiv\frac{1}{1-a(x) v}.
\end{equation}
Solving for $v$ yields
\begin{equation}
\label{eqsb11}
v=\frac{1\pm\sqrt{1-4a(x)}}{2a(x)},
\end{equation}
where, in order to ensure the correct value at steady state, the negative (positive) branch is appropriate for $v_\sts<2$ ($v_\sts>2$).

This theory of volume dynamics, in both its full and linearized version, shows very good agreement with a numerical solution of the full pump-leak model, as shown in Fig.~\ref{figSM1}b. Note that here, the volume follows the potential according to $v(\psi)=1/\left[1-\exp\left(e\psi/\kbt\right)\right]$ (dahsed black line) throughout the dynamics.

The pump-leak numerical scheme is similar to that of the RVD case with the exception that $p=0.02g$ as an initial value, and the regulatory dynamics are initiated by assuming a two-fold increase in $g_\Na$, while keeping $g_\K$ and $g_\Cl$ fixed. 

\end{widetext}
\end{document}